\documentstyle[revtex]{aps}
\begin{document}
\draft
\begin{title}
 Cosmology, Oscillating Physics and Oscillating Biology
\end{title}
\author{Pablo D. Sisterna \cite{aaa} and Hector Vucetich}
\begin{instit}
Departamento de Fisica, Facultad de Ciencias Exactas \\
Universidad Nacional de La Plata \\
C.C.67 1900 - La Plata, ARGENTINA
\end{instit}
\begin{abstract}
According to recent reports there is an excess correlation and an apparent
regularity in the galaxy one-dimensional polar distribution with a
characteristic scale of 128 $h^{-1}$ Mpc. This aparent spatial
periodicity can be naturally explained by a time oscillation of the
gravitational constant $G$. On the other hand, periodic growth features
of bivalve and coral fossiles appear to show a periodic component in the
time dependence of the number of days per year. In this letter we show that
a time oscillating gravitational constant with similar period and amplitude
can explain such a feature.
\end{abstract}

\pacs{90, 98, 98.80.-k, 98.80.dr}

Our  theoretical  description  of nature  presents  a  set  of  parameters,
the   so   called fundamental constants,  that  have  to  be  determined
from experience. It is generally thougth that this is so  because
we lack a unified  theory  of  all  interactions.  The  time
variation of any fundamental constant might be  one  of  the
few low energy phenomena  which  could  manifest  this  ``new
physics''. The time evolution of these parameters is supposed
to be governed by dynamics of cosmological  origin,  so  its
variation rate is likely to be of the same order  (or  less)
than the Hubble rate. We should seek then for low energy and
long term (millions of years)  phenomena \cite{mac,wolfe,shly}
or else for high precision measurements (ten  or  more  places  in  the
decimal expansion) of some ``constant''  parameter,  separated
by a few years \cite{dam,hel}. Actually,  superstring  theories \cite{schw}
and Kaluza-Klein theories \cite{kaluza} have cosmological solutions in  which
the low-energy fundamental constants vary with  time
\cite{yong,kei,moha,chodos,Marciano}.

Broadhurst, Ellis, Koo and Szalay \cite{Broad} combined data from four
distinct deep pencil-beam surveys at the north and south Galactic poles to
produce a well sampled distribution of galaxies by redshift on a linear
scale extending to 2000 $h^{-1}$ Mpc. They reported a periodicity in the
galaxy distribution of 128 $h^{-1}$ Mpc. Soon afterwards Morikawa
\cite{Morikawa} noted that this aparent spatial periodicity could be
naturally explained by a time oscillation of the Hubble parameter. In his
model the oscillation was produced by a massive scalar field non-minimally
coupled to gravity, inducing also the time oscillation of the gravitational
constant $G$. Hill, Steinhardt and Turner \cite{Hill} proposed different
scenarios, including a time oscillating Hubble parameter, but also
the oscillation of atomic lines as an
alternative explanation of the red-shift
galaxy distribution. An oscillation in the Rydberg constant due to the
variation of the fine structure
constant $\alpha$ or the electron mass $m_{e}$
requires a modification of the standard model in which $\alpha$ or $m_{e}$
become dependent on a scalar field. Thus an oscillating $\alpha$ introduces
a Yukawa potential between samples with non-zero electrostatic enegy
contribution, while an oscillating $m_{e}$ induces a similar interaction
now proportional to the lepton numbers of the samples \cite{Sudarsky}.
In this context both possibilities were shown to be ruled out \cite{Sudarsky}
by the experiment of Braginsky and Panov \cite{Braginskii}, leaving the
Hubble oscillating scenario as a solely candidate to analyze.

Even if the periodicity reported by
Broadhurst {\em et.al.} \cite{Broad} were
an artifact of small angle sampling of galaxy distribution, the problem of an
oscillating gravitational constant is interesting enough to deserve a
different observational analysis. There are indeed superstring models that
predict oscillating fundamental constants. These are functions of massive
dilaton fields associated with the size of extra dimensions \cite{yong,kei}.
To account a period $c\tau\simeq 128 h^{-1}$ Mpc the dilaton mass $m$ must be
$O(10^{-31} eV)$. In Ref.\ \cite{Hill} the time dependence proposed for the
gravitational constant is \cite{Hill}
\begin{equation}
	G=G_{0}+G_{1}\cos[2m(t-t_{0})+\psi],
\end{equation}
and the equation for the cosmological scale factor is
\begin{equation} \label{eq:cos}
       H^2=- \frac{\dot{G}}{G} H + \frac{8 \pi G \rho}{3},
\end{equation}
where $H$ is the Hubble parameter. From Eq.\ \ref{eq:cos} it is easy to see
that the leading correction to the Hubble parameter is
$H\simeq\bar{H} [1-\frac{1}{2}(\dot{G}/G \bar{H})]$.
Then, since $H\equiv \dot{a}/a$ and $1+z\propto a^{-1}$, one finds that
\begin{equation}
     \frac{dz}{dz_0}=1-\frac{\dot{G}}{2G\bar{H}}=
          1+{\cal A}\cos[2m(t-t_0)+\psi],
\end{equation}
where we have defined the amplitude ${\cal A}=-m G_1 /\bar{H}G_0$.
Following Ref.\ \cite{Hill}, if the Universe is spatialy homogeneous,
with a uniform galaxy density per comoving volume $n_0$, the number of
galaxies $dN$ in a solid angle $d\Omega$ with red shift between $z$ and
$z+dz$ is modulated compared to the distribution in the absence of
oscillations in the following way:
\begin{equation}
     \frac{dN}{z^2 dz d\Omega}=\left( \frac{dN}{z_0^2 dz_0 d\Omega}\right)
      \frac{dz_0}{dz}
\end{equation}
where to lowest order $z^2\simeq z_0^2$. We have then an apparent variation
in the density of galaxies which is isotropic and has peaks lying on
concentric spherical shells at periodically spaced radii. An amplitude
${\cal A}\simeq 0.5$ (or equivalently about 1\% for $G_1/G_0$),
a phase $\psi\simeq 0$ and a period $c\tau\simeq L=128 h^{-1}$
Mpc approximately reproduces the result of Ref.\ \cite{Broad}. Such a period
is similar to the time scales involved in the paleontological record.

  It is well known that several taxons record growth rhythms in their
skeletons; i.e. periodic markings locked to the astronomical cycles
of day, month and year \cite{Scrutton}. From these growth rhythms,
the number of days per year, days per month and months per year have
been obtained as functions of geological time
\cite{Scrutton,Lambeck,Lambeck2}. This parameters can be simply expressed in
terms of the Earth's rotation $\Omega$, the mean motion of the Sun
$n_{\odot}$ and the mean motion of the moon $n_{_)}$ as follows:
\begin{equation}
     N_{d/y}=\frac{\Omega}{n_{\odot}}-1,
\end{equation}
\begin{equation}
     N_{d/m}=\frac{\Omega-n_{\odot}}{n_{_)}-n_{\odot}},
\end{equation}
\begin{equation}
     N_{m/y}=\frac{n_{_)}-n_{\odot}}{n_{\odot}}.
\end{equation}
The effect of a varying $G$ on the Earth-Moon system can be studied under the
adiabatic hypothesis, as stated in references \cite{sv1,sv2}, so the Keplerian
equations of motion mantain their form, and $G$ is replaced by the appropiate
time function. Morover, in a reference system where $G$ depends only on time
but not on space, angular momentum is still conserved \cite{sv1}. Then we
obtain
\begin{equation}
     \frac{\dot{n}_{\odot}}{n_{\odot}}=2\frac{\dot{G}}{G},
\end{equation}
\begin{equation}
     \frac{\dot{n}_{_)}}{n_{_)}}=2\frac{\dot{G}}{G}+
                \left.\frac{\dot{n}_{_)}}{n_{_)}}\right| _t,
\end{equation}
\begin{equation}
     \frac{\dot{\Omega}}{\Omega}=-\gamma\left.\frac{\dot{n}_{_)}}{n_{_)}}
      \right| _t+(2-\beta)\frac{\dot{G}}{G},
\end{equation}
where $\beta\simeq 1.83$ depends on the mass and pressure distribution
in the Earth's interior \cite{sv1}, $\gamma\simeq 1.84$ is related to the
tidal couple of the Earth-Moon system \cite{Lambeck},
$\dot{n}_{_)}|_{t}$ is the tidal acceleration of the lunar longitude
and $-(2-\beta)\frac{\dot{G}}{G}$ is the change of the moment of inertia
due to the change of $G$ \cite{sv1}.
 According to Lambeck\cite{Lambeck2} the palaeontological data suggest that
there may be significant systematic errors in the counts, and it is
convenient to introduce parameters $\Delta n_{\odot}$,$\Delta n_{_)}$
and$\Delta\Omega$ accounting for such errors in the estimation of the
planetary angular velocities. Then for a time oscillating $G$ the final
expressions for the observables are
\begin{equation}
	\frac{N_{d/y}}{N^{0}_{d/y}}-1=\beta \frac{G_{1}}{G_{0}}
          [\sin(\omega\tilde{t}-\phi)-\sin(\phi)]-
          \left.\frac{\dot{\Omega}}{\Omega}\right| _{t}\tilde{t}-
          \frac{\Delta\Omega}{\Omega}+\frac{\Delta n_{\odot}}{n_{\odot}},
\end{equation}
\begin{equation}
          \frac{N_{m/y}}{N^{0}_{m/y}}-1=\beta \frac{G_{1}}{G_{0}}
          [\sin(\omega \tilde{t}-\phi)-\sin(\phi)]+
          \left.\left(\frac{\dot{n_{_)}}}{n_{_)}}\right| _{t}-
          \left.\frac{\dot{\Omega}}{\Omega}\right| _{t}\right)\tilde{t}-
          \frac{\Delta\Omega}{\Omega}+\frac{\Delta n_{_)}}{n_{_)}},
\end{equation}
\begin{equation}
          \frac{N_{d/m}}{N^{0}_{d/m}}-1=
         -\left.\frac{\dot{n_{_)}}}{n_{_)}}\right| _{t}\tilde{t}+
          \frac{\Delta n_{_)}}{n_{_)}}+
          \frac{\Delta n_{\odot}}{n_{\odot}},
\end{equation}
where $\tilde{t}$ is minus the geological (not ephemeris) time \cite{sv1} and
$\phi=\frac{\pi}{2}-\psi$.

We used the data carefully filtered by Lambeck\cite{Lambeck2} according to
biological reliability
criteria \cite{Lambeck}. These data (37 points) are shown in
Table~\ref{tab:ldata}. Originally Lambeck adjusted the data assuming
that only (constant) dissipation mechanisms are the responsible for
$\dot{\Omega}$ and $\dot{n}_{_)}$. The adjusted values of
$\dot{\Omega}$ and $\dot{n}_{_)}$ were in good agreement with the modern
astronomical values. The residuals showed no obvious
systematic trends, indicating that the growth rings present a high degree
of confidence.
In our case we adjusted simultaneously the three curves with and
without the oscillatory parameters (i.e. with and without an oscillatory $G$),
using the Levenberg-Marquardt least-squares method. The $\chi^{2}$
significances of the adjustments are shown in Table~\ref{tab:significances}.
There is still a debate on the correct values for standard deviations of
the palaeontological number counts, so the absolute value of the $\chi^{2}$
significance is not important. What it is really important is the {\it change}
of the significance when we include the oscillatory parameters. We see that
there is a conspicuous increase of significance when we include
the oscillatory parameters. For this last adjustment the best fit values are
 shown in Table~\ref{tab:results}, together with independent $95\%$
confidence limits. The upper confidence limit of $G_{1}/G_{0}$ is about
0.009, which is the desired amplitude accounting for the red-shift survey,
i.e.,we have marginal consistency with the oscillating $G$ hypothesis. The
best fit value of the period of oscillation remarkably coincides with the
galaxy distribution period, although the confidence limits weaken its
relevance. The values for (tidal)
$\dot{n}_{_)}/n_{_)}$,$\dot{\Omega}/\Omega$ and $\dot{n}_{\odot}/n_{\odot}$
agree with other estimates \cite{Lambeck2}. The phase $\psi$ is consistent
with the zero value proposed in \cite{Hill}, which implies a zero value for
the present rate of change of $G$. Our fit is then consistent with the
current upper bounds on the time variation of $G$ based upon the Viking
radar-echo experiments \cite{hel} [note that only upper bounds based on
present observations are valid if $G$ oscillates, so the much more stringent
upper bounds of \cite{sv1,sv2} do not apply because they are also based in
long-term (several oscillation periods) phenomena].  In order to test the
sensibility of the solution, we made the same adjustments using original
data from several sources, as shown in Table~\ref{tab:data}, including
bivalves, corals, cephalopods, brachiopods and estromatolites and totalizing
61 points.  As shown in Table~\ref{tab:significances} in this case both
significances are small, showing that the filtering of data as made by
Lambeck introduces a bias towards the oscillatory hypothesis.

We conclude that our results do not exclude an oscillating gravitational
constant inducing a periodic galaxy distribution. Indeed the significance
of the Lambeck data adjustment suggests thet there is an oscillatory
component in the time evolution of planetary orbit ratios. However,
there are several uncertainties in our model that forbid a definite
conclusion. In the first place, paleontological growth rhythms are
subject to large variations, and should be handled with care
\cite{Lutz,Hughes,Rosenberg}.
Secondly, it should be noted that the changes in the resonance structure of
the oceans due to continental drift provoke considerable variations of the
Earth-Moon tidal torque within 100 million years time scales \cite{Brosche}.
This fact could account as well for the time oscillation of the number of
days per year. As a result we can only state that an upper bound for the
oscillation amplitude of the gravitational constant is $G_{1}/G_{0}<0.01$
(taken from the upper confidence limit of Table~\ref{tab:results}).
The Hubble oscillating hypothesis is strongly testable beacause it predicts
the same periodic patterns in all directions. Also, as stated by Morikawa
\cite{Morikawa} both the density contrast and the distance to the
nearest peak is not clear from the survey. Then it is worth making further
deep-pencil surveys in the same and other directions in the sky.
Summarizing, the coincidences which we observe in this work (coincidences
may become consequences \cite{Barrow}) are significant enough so as to
become a subject of further numerical and observational research.

\pagebreak
\mediumtext
\begin{table}
\caption{Lambeck selection of counting and geological data.\label{tab:ldata}}
\begin{tabular}{rllll}
Geologic time [M.Y.] &Counting &Adopted S.D. &Organisms &Data type  \\
\tableline
-300 & 385   & 4.2  & corals   & da   \\
-320 & 398   & 6    & corals   & da   \\
-370 & 398   & 1.7  & corals   & da   \\
-420 & 400   & 6    & corals   & da   \\
-440 & 412   & 6    & corals   & da   \\
   0 & 360   & 4    & corals   & da   \\
-370 & 401   & 6    & corals   & da   \\
-370 & 30.66 & 0.5  & corals   & dm   \\
-330 & 30.2  & 0.4  & corals   & dm   \\
   0 & 359   & 2.1  & bivalves & da   \\
 -70 & 375   & 3.5  & bivalves & da   \\
-220 & 372   & 3.5  & bivalves & da   \\
-290 & 383   & 3.5  & bivalves & da   \\
-340 & 398   & 4.2  & bivalves & da   \\
-360 & 406   & 6    & bivalves & da   \\
   0 & 29.2  & 0.4  & bivalves & dm   \\
 -14 & 29.4  & 0.5  & bivalves & dm   \\
 -38 & 29.8  & 0.8  & bivalves & dm   \\
 -54 & 29.6  & 0.6  & bivalves & dm   \\
 -70 & 29.9  & 0.4  & bivalves & dm   \\
-220 & 29.7  & 0.6  & bivalves & dm   \\
-300 & 30.2  & 0.6  & bivalves & dm   \\
-350 & 30.4  & 0.6  & bivalves & dm   \\
-445 & 30.3  & 0.8  & bivalves & dm   \\
  -1 & 14.75 & 0.14 & bivalves & dm/2 \\
  -4 & 14.83 & 0.16 & bivalves & dm/2 \\
 -30 & 14.82 & 0.09 & bivalves & dm/2 \\
 -48 & 14.87 & 0.27 & bivalves & dm/2 \\
 -60 & 14.82 & 0.17 & bivalves & dm/2 \\
-100 & 14.88 & 0.12 & bivalves & dm/2 \\
-160 & 14.90 & 0.18 & bivalves & dm/2 \\
-230 & 14.91 & 0.07 & bivalves & dm/2 \\
-310 & 15.09 & 0.17 & bivalves & dm/2 \\
-370 & 15.25 & 0.60 & bivalves & dm/2 \\
   0 & 12.3  & 0.17 & bivalves & ma   \\
 -70 & 12.6  & 0.26 & bivalves & ma   \\
-220 & 12.6  & 0.26 & bivalves & ma   \\
\end{tabular}
\end{table}
\begin{table}
\caption{Original counting and geological data.\label{tab:data}}
\begin{tabular}{rlllll}
Geologic time [M.Y.] &Counting &Adopted S.D. &Organisms &Data type&Reference\\
\tableline
-300 & 385    & 4.3  & corals         & dy & \cite{Wells}      \\
-330 & 398    & 6.1  & corals         & dy & \cite{Wells}      \\
-390 & 397.6  & 1.8  & corals         & dy & \cite{Wells}      \\
-425 & 400    & 6.1  & corals         & dy & \cite{Wells}      \\
-465 & 412    & 6.1  & corals         & dy & \cite{Wells}      \\
   0 & 360    & 3.1  & corals         & dy & \cite{Wells}      \\
-390 & 401    & 6.1  & corals         & dy & \cite{Scrutton64} \\
-395 & 410    & 4.5  & corals         & dy & \cite{Mazzulo}    \\
-440 & 421    & 6.4  & corals         & dy & \cite{Mazzulo}3   \\
-390 &  30.63 & 0.11 & corals         & dm & \cite{Scrutton64} \\
-335 &  30.2  & 0.15 & corals         & dm & \cite{Johnson}    \\
-480 &  30.7  & 0.23 & corals         & dm & \cite{Panella}    \\
-395 &  31.5  & 0.45 & corals         & dm & \cite{Mazzulo}    \\
-440 &  32.4  & 0.64 & corals         & dm & \cite{Mazzulo}    \\
-390 &  13    & 0.29 & corals         & ma & \cite{Scrutton64} \\
-395 &  13    & 0.2  & corals         & ma & \cite{Mazzulo}    \\
-440 &  13    & 0.29 & corals         & ma & \cite{Mazzulo}    \\
   0 & 359.3  & 1.6  & bivalves       & da & \cite{Panella}    \\
 -70 & 375    & 2.6  & bivalves       & da & \cite{Panella}    \\
-220 & 371.6  & 3.6  & bivalves       & da & \cite{Panella}    \\
-290 & 383    & 2.3  & bivalves       & da & \cite{Panella}    \\
-350 & 398    & 1.8  & bivalves       & da & \cite{Panella}    \\
-390 & 405.5  & 6.0  & bivalves       & da & \cite{Panella}    \\
   0 &  29.22 & 0.08 & bivalves       & dm & \cite{Panella}    \\
 -10 &  29.52 & 0.08 & bivalves       & dm & \cite{Panella}    \\
 -22 &  29.42 & 0.15 & bivalves       & dm & \cite{Panella}    \\
 -51 &  29.7  & 0.09 & bivalves       & dm & \cite{Panella}    \\
 -70 &  29.85 & 0.10 & bivalves       & dm & \cite{Panella}    \\
-220 &  29.66 & 0.15 & bivalves       & md & \cite{Panella}    \\
-290 &  30.16 & 0.09 & bivalves       & dm & \cite{Panella}    \\
-350 &  30.37 & 0.11 & bivalves       & dm & \cite{Panella}    \\
-390 &  30.35 & 0.17 & bivalves       & dm & \cite{Panella}    \\
   0 &  12.35 & 0.07 & bivalves       & ma & \cite{Panella}    \\
 -70 &  12.64 & 0.11 & bivalves       & ma & \cite{Panella}    \\
-220 &  12.56 & 0.08 & bivalves       & ma & \cite{Panella}    \\
  -1 &  29.5  & 0.08 & bivalves       & dm & \cite{Berry75}    \\
  -5 &  29.66 & 0.15 & bivalves       & dm & \cite{Berry75}    \\
 -30 &  29.52 & 0.08 & bivalves       & dm & \cite{Berry75}    \\
 -45 &  29.72 & 0.28 & bivalves       & dm & \cite{Berry75}    \\
 -60 &  29.7  & 0.16 & bivalves       & dm & \cite{Berry75}    \\
-100 &  29.72 & 0.05 & bivalves       & dm & \cite{Berry75}    \\
-170 &  29.84 & 0.3  & bivalves       & dm & \cite{Berry75}    \\
-220 &  29.76 & 0.3  & bivalves       & dm & \cite{Berry75}    \\
-325 &  30.13 & 0.13 & bivalves       & dm & \cite{Berry75}    \\
-400 &  30.5  & 0.4  & bivalves       & dm & \cite{Berry75}    \\
 -70 &  29.65 & 0.18 & bivalves       & dm & \cite{Berry68}    \\
 -70 &  12.49 & 0.02 & bivalves       & ma & \cite{Berry68}    \\
-320 &  30.22 & 0.40 & cephalopods    & dm & \cite{Panella}    \\
-410 &  29.84 & 0.23 & cephalopods    & dm & \cite{Panella}    \\
-310 &  30.11 & 0.35 & cephalopods    & dm & \cite{Pompea}     \\
-395 & 407.75 & 3.2  & brachiopods    & da & \cite{Mazzulo}    \\
-425 & 419    & 4.5  & brachiopods    & da & \cite{Mazzulo}    \\
-395 &  31.38 & 0.32 & brachiopods    & dm & \cite{Mazzulo}    \\
-425 &  31.5  & 0.45 & brachiopods    & dm & \cite{Mazzulo}    \\
-395 &  13    & 0.15 & brachiopods    & ma & \cite{Mazzulo}    \\
-425 &  13    & 0.2  & brachiopods    & ma & \cite{Mazzulo}    \\
-540 & 424    & 6.4  & estromatolites & da & \cite{McGuhan}    \\
-150 &  30    & 0.64 & estromatolites & dm & \cite{Panella}    \\
-510 &  31.56 & 0.75 & estromatolites & dm & \cite{Panella}    \\
-510 &  33    & 0.64 & estromatolites & dm & \cite{Panella}    \\
-510 &  13    & 0.29 & estromatolites & ma & \cite{Panella}    \\
\end{tabular}
\end{table}
\narrowtext
\begin{table}
\caption{Significance of the $\chi^{2}$ test for all adjustments.
         \label{tab:significances}}
\begin{tabular}{lll}
Data source & number of adjusted & $\chi^{2}$ significance \\
            & parameters         &                         \\
\tableline
Original & 5 & below 10\% \\
Original & 8 & below 10\% \\
Lambeck  & 5 & 17\%       \\
Lambeck  & 8 & 85\%       \\
\end{tabular}
\end{table}
\begin{table}
\caption{Adjusted parameters of the curves fitting Lambeck data, together
with $95\%$ confidence limits.\label{tab:results}}
\begin{tabular}{lccc}
Parameter & best fit value & \multicolumn{2}{c}{confidence limits} \\
          &                & lower & upper                         \\
\tableline
$\Delta n_{\odot}$    & -1.3$\times 10^{-2}$ & -2.5$\times 10^{-2}$ &
1.6$\times 10^{-3}$ \\
$\Delta n_{_)}$       & 9.3$\times 10^{-3}$  & -4.1$\times 10^{-2}$ &
5.9$\times 10^{-2}$ \\
$\Delta\Omega$        & 1.9$\times 10^{-3}$  & -4.1$\times 10^{-3}$ &
7.9$\times 10^{-3}$ \\
$\frac{\dot{n}}{n}[10^{-11}$yr$^{-1}$            & 16.4 & 12 & 20  \\
$\frac{\dot{\Omega}}{\Omega}[10^{-11}$yr$^{-1}$  & 32.0 & 29 & 35  \\
$\frac{G_{1}}{G_{0}}$ & 6.5 $\times 10^{-3}$     & 3.8$\times 10^{-3}$ &
9.2$\times 10^{-3}$ \\
period [$10^{6}$yr]   & 486        & 370      & 680   \\
$\psi$                & 0.27       & -0.53    & 0.83  \\
\end{tabular}
\end{table}

\begin{thebibliography}{99}
%
\bibitem[*]{aaa} also at Departamento de F\'{i}sica,
  F.C.E.y N, Universidad Nacional de Mar del Plata,
  Funes 3350 - (7600) Mar del Plata, ARGENTINA
%
\bibitem{mac} M. W. Mc.Elhinny, S. R. Taylor and D. J. Stevenson,
Nature (London) {\bf 271}, 316 (1978).
%
\bibitem{wolfe} A. M. Wolfe, R. L. Brown and M. S. Roberts,
Phys.\ Rev.\ Lett. {\bf 37}, 179 (1976).
%
\bibitem{shly} A. I. Shlyakhter, Atomki  Report A/I (1983);
Nature (London) {\bf 264} , 340 (1976).
%
\bibitem{dam} T. Damour, G. W. Gibbons and J. H. Taylor,
Phys.\ Rev.\ Lett. {\bf 61}, 1151 (1988).
%
\bibitem{hel} R. H. Hellings, P. J. Adams, J. D. Anderson, M. S. Keesey,
E. L. Lau, E. M. Standish, V. M. Canuto and I. Goldman,
Phys.\ Rev.\ Lett. {\bf 51}, 1609 (1983).
%
\bibitem{schw} J. H. Schwartz, Phys. Rep. {\bf 89},
 223 (1982); M. B. Green J. H. Schwartz, Phys.\ Lett.{\bf 199B}, 117 (1984).
%
\bibitem{kaluza} T. Kaluza, Sitz.\ Preuss.\ Akad.\ Wiss.\ Phys.\ Math.
{\bf K1}, 966 (1921); O. Z. Klein, Phys. {\bf 37}, 895 (1926).
%
\bibitem{yong} Y. Wu and  Z. Wang, Phys.\ Rev.\ Lett.
{\bf 57}, 1978 (1986).
%
\bibitem{kei} K. Maeda, Mod.\ Phys.\ Lett. {\bf A3}, 243 (1988);
%
\bibitem{moha} S. M. Barr and P. K. Mohapatra,
Phys.\ Rev. {\bf D38},3011 (1988).
%
\bibitem{chodos} A. Chodos and S. Detweiler, Phys.\ Rev. {\bf D21},
2167 (1980). M. Gleiser and J. G. Taylor,
Phys.\ Rev. {\bf D31}, 1904 (1985).
%
\bibitem{Marciano} W. J. Marciano, Phys.\ Rev.\ Lett. {\bf 52}, 489 (1984).
%
\bibitem{Broad}T. J. Broadhurst, R. S. Ellis, D. C. Koo and
A. S. Szalay, Nature (London) {\bf 343}, 726 (1990).
%
\bibitem{Morikawa} M.Morikawa, Astrophys.\ J. {\bf 369}, 20 (1991).
%
\bibitem{Hill} C. T. Hill, P. J. Steinhardt and M. S. Turner,
Fermi National Accelerator Laboratory prepint 90/129-T (1990).
%
\bibitem{Sudarsky} D.Sudarsky, Fermi Institute preprint, 1992.
%
\bibitem{Braginskii} V.B.Braginskii and V.I.Panov, Zh.\ Eksp.\ Teor.\ Fiz.\
{\bf 61}, 873 (1971).
%
\bibitem{Scrutton} C. T. Scrutton, in {\it Tidal Friction and the
Earth's Rotation} (eds Brosche, P.  Sundermann, J.) 154-196
(Springer, New York, 1978).
%
\bibitem{Lambeck} K. Lambeck, {\it The Earth's Variable Rotation}
(Cambridge University Press, New York, 1980).
%
\bibitem{Lambeck2} K. Lambeck, in {\it Tidal Friction and the Earth's
Rotation}, ed. by P. Brosche and J. Sundermann,
(Springer-Verlag, Berlin 1978).
%
\bibitem{sv1} P. D. Sisterna and H. Vucetich, Phys.\ Rev. {\bf D41},
1034 (1990).
%
\bibitem{sv2} P. D. Sisterna and H. Vucetich, Phys.\ Rev. {\bf D44},
3096 (1991).
%
\bibitem{Lutz} R. A. Lutz and D. C. Rhoads, Science {\bf 198}, 1222 (1977).
%
\bibitem{Hughes} W. W. Hughes, Geophysical Surveys {\bf 7}, 169 (1985).
%
\bibitem{Rosenberg} G. D. Rosenberg, Geophysical Surveys {\bf 7}, 185 (1985).
%
\bibitem{Brosche} P. Brosche, in {\it Sun and Planetary System},
ed. by W. Fricke and G. Teleki (Reidel, D. Publishing Company).
%
\bibitem{Barrow} J.D.Barrow and F.J.Tipler, {\it The Anthropic Cosmological
 Principle} (Oxford University Press, Oxford, 1988).
%
\bibitem{Wells} J.W.Wells, Nature (London) {\bf 197}, 948 (1963).
      J.W.Wells, in {\it Palaeogeophysics} ed. by S.K.Runcorn
         (London, Academic Press, 1970).
%
\bibitem{Scrutton64} C.T.Scrutton, Palaeontology {\bf 7}, 552 (1965).
%
\bibitem{Mazzulo} S.J.Mazzulo, Bull.Geol.Soc.Am. {\bf 82}, 1085 (1971).
%
\bibitem{Johnson} G.A.L. Johnson and J.R.Nudds, in {\it Growth Rhythms and
the History of the Earth's Rotation}, ed. by G.D.Rosenberg and S.K.Runcorn
(London, Wiley, 1975).
%
\bibitem{Panella} G.Panella, Astrophys.Space Sci. {\bf 16}, 212 (1972).
%
\bibitem{Berry75} W.B.N.Berry and R.M.Baker, in {\it Growth Rhythms and the
History of the Earth's Rotation}, ed. by G.D.Rosenberg and S.K.Runcorn
(London, Wiley, 1975).
%
\bibitem{Berry68} W.B.N.Berry and R.M.Baker, Nature (London) {\bf 217}, 938
(1968).
%
\bibitem{McGuhan} A.McGuhan, Abstr.Mtg.Geol.Soc.Am. {\bf 1967}, 145 (1967).
%
\bibitem{Pompea} S.M.Pompea, P.G.K.Kahn and R.B.Culver, Vistas in Astronomy
1979 {\bf 23}, 185 (1979).
%
\end{thebibliography}
\end{document}